\documentclass[prl,twocolumn,noshowkeys,superscriptaddress,preprintnumbers,amsmath,amssymb, showpacs]{revtex4}

\newcommand{\abs}[1]{\left\vert #1 \right\vert}

\newcommand{\Tr}[1]{\textrm{Tr}\left( #1 \right)}
\newcommand{\erf}[1]{\textrm{erf}\left( #1 \right)}

\usepackage{amsmath}
\usepackage{bbm,amsfonts}

\usepackage{amsthm}
\theoremstyle{definition}

\input{Qcircuit.tex}
\usepackage{graphicx}

\newcommand{\DE}[1]{\left\{ #1 \right\}}

\newcommand{\esp}[1]{{\mathsf{#1}}}

\begin{document}

\title{General theory of measurement with two copies of a quantum state}

\author{Ariel Bendersky}
\author{Juan Pablo Paz}
\affiliation{Departamento de F\'\i sica,``Juan Jos\'e Giambiagi'', FCEyN UBA, Pabell\'on 1, Ciudad Universitaria, 1428 Buenos Aires, Argentina}
\author{Marcelo Terra Cunha}
\affiliation{Departamento de Matem\'atica, Universidade Federal de Minas Gerais, Caixa Postal 702, Belo Horizonte, 30123-970, Brazil} 
\date{\today}

\begin{abstract} 
We analyze the possible results of the most general measurement on two copies of a quantum state. We show that $\mu$ can label a set of outcomes of such measurement if and only if there is a family of completely co--positive (ccP) maps $C_\mu$ such that the probability of occurrence $Prob(\mu)$ is the fidelity of the map $C_\mu$, i.e. $Prob(\mu)= Tr(\rho\ C_\mu(\rho))$ which must add up to the fully  depolarizing map. This implies that a POVM on two copies induces a measure on the set of ccP maps (i.e., a ccPMVM). We present examples of ccPMVM's and discuss their tomographic applications showing that two copies of a state provide an exponential improvement in the efficiency of quantum state tomography. This enables the existence of an efficient universal detector. 
\end{abstract}


\maketitle


One of the postulates of quantum theory tells us how to compute probabilities for the outcomes of measurements: If the system was prepared in the state $\rho$, for every outcome $\mu$ of a measurement there is a projector $P_\mu$ such that the probability of occurrence of $\mu$ is the expectation value of the projector $P_\mu$ in the state $\rho$, i.e. $Prob(\mu)=Tr(\rho P_\mu)$. To represent mutually exclusive outcomes the projectors must be orthogonal and they must add up to the identity to ensure that the total probability is unity (i.e. $P_\mu P_\nu = \delta_{\mu\nu} P_\mu$, $\sum_\mu P_\mu= I$). This postulate, originally formalized by von Neumann \cite{vonNeumann}, was extended in the 1970's \cite{Kraus} when the notion of generalized measurement was introduced. In such measurement a positive operator $A_\mu$ (not necessarily a projector) is associated with every outcome $\mu$ and the probability of occurrence of $\mu$ is  $Prob(\mu)=Tr(\rho A_\mu)$. The operators $A_\mu$ add up to the identity and define a so-called positive operator valued measure (POVM). Neumark's Theorem \cite{Peres} establishes that POVM measurements are equivalent to projective measurements for an extended system: Any POVM can be implemented via a projective measurement on the original system supplemented with an appropriately chosen ancillary system. 

In this paper we analyze the predictions of quantum theory concerning the results of measurements performed when two identically prepared quantum systems are simultaneously available. More precisely, we assume that a source produces the state $\rho^{(A,B)}=\rho\otimes\rho$ ($A$ and $B$ label two systems prepared in the same state $\rho$). Our goal is twofold: a) To determine the possible distributions for the measurement outcomes; b) To present a solution to the problem of efficient universal state tomography using copies. We divide the presentation in two parts. First we prove the following Theorem (ccPMVM): Given two systems prepared in the same state $\rho$ then $\mu$ can label a set of possible outcomes of a measurement on $\rho^{(A,B)}=\rho\otimes\rho$ if and only if there is a family of completely co--positive (ccP) maps $C_\mu$ such that the probability of occurrence $Prob(\mu)$ is the fidelity of the map $C_\mu$, i.e. $Prob(\mu)= Tr(\rho\ C_\mu(\rho))$. The maps satisfy the condition $\sum_\mu C_\mu={\mathcal E}$ where ${\mathcal E}$ is the map for which ${\mathcal E}(\rho)=I$ for any state $\rho$. Moreover $C_\mu$ must be ccP which means that the composition of $C_\mu$ with the transposition must be completely positive. This Theorem establishes an interesting connection between families of ccP maps and general measurements with copies. As a consequence, a measurement with copies defines a ``ccP map valued measure" (ccPMVM) and vice versa. In the second part of this Letter we establish the tomographic power of this type of measurement showing that availability of two copies gives an exponential advantage in solving the general problem of universal quantum state tomography enabling us to construct an efficient universal quantum state detector and to efficiently estimate partial purities, concurrences, and other interesting quantities. 

To prove the ccPMVM Theorem it is crucial to use the so-called Jamio\l kowski isomorphism\cite{Jami} that establishes a one to one correspondence between linear operators on the space $\esp{H}\otimes\esp{H}$ and linear super-operators on the space $\esp{H}$ (a super-operator on $\esp{H}$ is a map on the space of operators over $\esp{H}$). For every super-operator $\tilde C$ on $\esp{H}$ we can associate an operator $\hat C$ on $\esp{H}\otimes\esp{H}$ and vice versa. This one to one correspondence is realized by the following identity: $\hat C=(\tilde C\otimes I)(\ket{\mathcal I}\bra{\mathcal I})$ where $\ket{\mathcal I}$ the unnormalized maximally entangled state $\ket{\mathcal I}=\sum_i\ket{ii}$. The isomorphism relates positive Hermitian operators on $\esp{H}\otimes\esp{H}$ with completely positive Hermitian super-operators on $\esp{H}$. In particular, the identity operator is associated to the completely depolarizing super-operator $\mathcal E$ for which the image of every trace one operator is the identity. Using this isomorphism the ccPMVM Theorem can be proved as follows: An outcome $\mu$ of a generalized measurement on two copies prepared on the state $\rho^{(A,B)}=\rho\otimes\rho$ is characterized by a positive operator $\hat C_\mu$. The probability of such outcome is $Prob(\mu)=Tr(\rho\otimes\rho\ \hat C_\mu)$. Jamio\l kowski isomorphism ensures that for every positive operator $\hat C_\mu$ there is a completely positive super-operator $\tilde C_\mu$ such that $Prob(\mu)=Tr(\rho\otimes\rho\ (\tilde C_\mu\otimes I)(\ket{\mathcal I}\bra{\mathcal I}))$. By replacing the explicit form of the state $\ket{\mathcal I}$ the trace over the second copy can be computed and the probability $Prob(\mu)$ can be rewritten as $Prob(\mu)=Tr(\rho\ \tilde C_\mu(\rho^T))$, where  $\rho^T$ denotes the usual transposition of $\rho$. Therefore, the probability of every outcome of a generalized measurement is $Prob(\mu)=Tr(\rho\ C_\mu(\rho))$ where $C_\mu = \tilde{C}_{\mu} \circ {\mathcal{T}}$, with ${\mathcal{T}}$ denoting the transposition as a map. Jamio\l kowski isomorphism ensures that $\tilde C_\mu$ is completely positive, which implies that $C_\mu$ is completely co-positive. Moreover, as the POVM operators $\hat C_\mu$ add up to the identity, the corresponding super-operators $\tilde C_\mu$ must add up to the completely depolarizing map. The relation ${\mathcal{E}}\circ {\mathcal{T}} = {\mathcal{T}}\circ {\mathcal{E}} = {\mathcal{E}}$ completes the prove of the ccPMVM Theorem. 

This Theorem characterizes all measurements with two copies and shows that the fidelity of certain families of positive super-operators has a direct physical meaning as it can always be realized as the probability of a generalized measurement with copies. We now analyze an example which is significant from the point of view of quantum state tomography. Suppose that we have two copies of the  state $\rho$ of an $n$--qubit system: $\rho^{(A,B)}=\rho\otimes\rho$. We will assume that we perform a Bell measurement on all pairs formed by the $j$--th qubit of each copy as shown in Figure 1. We write the state of each copy as $\rho=\sum_{q,p} c_{q,p} T(q,p) /N$ where $q=(q_1,\ldots,q_n)$ and $p=(p_1,\ldots, p_n)$ are binary $n$--tuples. The generalized Pauli operators $T(q,p)$ are $n$--fold tensor products of the identity and the three Pauli operators on each qubit: $T(q,p)=X^{q_1}Z^{p_1}\otimes\ldots \otimes X^{q_n}Z^{p_n}  (i)^{q p}$ (here $q p=\sum_k q_kp_k$). Real coefficients $c_{q,p}$ are such that $c_{q,p} =Tr(\rho\ T(q,p))$. 
\begin{figure}
 \includegraphics[width=35mm]{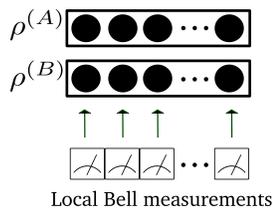}
\caption{Proposed scheme for full state tomography.\label{figEsquema}}
\end{figure} 

We can collect the outcomes of all the Bell measurements in two binary $n$--tuples $(a,b)$ where $(a_k,b_k)$ identify the state $|\beta_{a_k,b_k}\rangle$ detected at site $k=1,\ldots ,n$ (Bell state $|\beta_{a_k,b_k}\rangle$ is an eigenstate of $X_k\otimes X_k$ and $Z_k\otimes Z_k$ with eigenvalue $(-1)^{a_k}$ and $(-1)^{b_k}$ respectively). After some manipulation, we can show that the probability of occurrence for every possible outcome $Prob(a,b)$ is  $Prob(a,b)=\sum_{q,p} (-1)^{a q+b p +q p}c_{q,p}^2 /N^2$. These probabilities are, as the ccPMVM Theorem ensures, the fidelities of ccP maps. Indeed, one can show that $Prob(a,b) =Tr(\rho\ C_{a,b} (\rho))$ where the corresponding map is $C_{a,b}(\rho)=T(b,a) \rho^T T(b,a)/N$. The ccP character of these maps is evident since they are obtained as the composition of the transposition with a completely positive super-operator. It is useful to analyze the simplest case of a single qubit where the coefficients $c_{1,0}$, $c_{0,1}$  and $c_{1,1}$ are nothing but the three Cartesian components of the Bloch vector $\vec p$ parametrizing the state as a linear combination of the three Pauli's:  $\rho=(I+\vec p \cdot \vec \sigma)/2$.  Then, the maps $C_{a,b}$ are such that  $C_{a,b}(\rho)=(I+\vec p_{a,b} \cdot \vec\sigma)/4$. These operators are proportional to states with polarization vectors $\vec p_{a,b}=(-1)^a p_x \hat x+ (-1)^{a+b+1} p_y \hat y+ (-1)^b p_z \hat z$. Therefore, the map $C_{1,1}$, corresponding to the singlet $|\beta_{1,1}\rangle$, realizes a full inversion on the Bloch sphere. The other Bell states have maps corresponding to  reflections about the three Cartesian planes (where one Cartesian component of $\vec p$ changes sign\footnote{Those maps are well known examples of positive, but not completely positive maps and it is very interesting that they naturally appear in this context.}). Adding these four maps we obtain the fully depolarizing one. Probabilities for the four Bell measurements are quadratic in the components of $\vec p$: $Prob(a,b)=(1+\vec p\ \cdot \vec p_{a,b})/4$. 

Interestingly, the generalized measurement with two copies described above can be used to device an efficient strategy for quantum state tomography (QST)
\cite{citeulike:541803, 2004PhRvA..70e2321R} a task whose goal is to extract complete information about the state $\rho$. This usually requires performing an exponentially large number of different experiments on identically prepared systems. As $\rho=\sum_{q,p} c_{q,p} T(q,p)/N$ is a linear combination of $N^2$ generalized Pauli operators $T(q,p)$, a straightforward way to determine $c_{q,p}$ is to measure all Pauli's on each qubit. This requires $N^2$ different experimental setups and an exponentially expensive post-processing (estimation of each coefficient $c_{q,p}$ typically requires the use of a different experimental setup). However, we will show that if two copies are available the complexity of QST is drastically reduced.  The use of copies was analyzed before in this context and it was shown that it enables us to extract more information than single copy measurements \cite{1999PhRvA..60..126V, 1995PhRvL..74.1259M, 1998PhRvL..81.1351L, 1998PhRvL..80.1571D, 2004PhRvL..93w0501C, 2006PhRvL..97e0501A}. Here we generalize these results showing that the number of experiments required to gather the information to estimate every coefficient $c_\alpha^2$ with fixed precision is independent of the number of qubits. A possible strategy to determine all $c_{q,p}^2$ is to estimate all probabilities $Prob(a,b)$ and later to obtain such coefficients by performing a Hadamard transform. However, this would be highly inefficient method since to obtain such coefficients we would need to estimate an exponentially large number of exponentially small probabilities (it is simple to see that probabilities obey $Prob(a,b)\lesssim 1/N$). On the contrary, to efficiently estimate $c_{q,p}^2$ we can proceed as follows: After performing  the Bell measurements described in Figure 1 we can multiply the detected values of the operators $X_j \otimes X_j$ and $Z_j\otimes Z_j$ to obtain the value of any Pauli operator of the form $T(q,p) \otimes T(q,p)$. By repeating this
  experiment a number of times we can estimate the corresponding expectation value of such operator, which is precisely what we need to compute $|c_{q,p}|$ since:
\begin{equation}
\left\langle T(q,p) \otimes T(q,p) \right\rangle_{\rho\otimes\rho} 
= c_{q,p}^2.
\end{equation}
We can show that the number of experimental runs, $M_E$, required to estimate any $c_{q,p} $ (up to a sign) with fixed precision is independent of the number of qubits and is only fixed by the precision. In fact, every measurement yields binary values for $T(q,p) \otimes T(q,p) $. Therefore, after $\tilde M_E$ repetitions we compute the average result that we denote $\tilde c_{q,p}^2$. The central limit theorem implies that the standard deviation $\sigma_{q,p}$ for $c_{q,p}^2$ satisfies $\sigma_{q,p} \leq {1}/{\sqrt{\tilde M_E}}$. Thus, there is a number $k$ such that $c_{q,p}^2 \in \left[ \tilde c_{q,p}^2-k\sigma_{q,p}; \tilde c_{q,p} ^2+k\sigma_{q,p} \right]$ with probability $p$.  This bound propagates to $|c_{q,p}|$ that with the same probability $p$ will be found in an interval centered at $|\tilde c_{q,p}|$ with a width $\displaystyle{\frac{k\sigma_{q,p}}{2\abs{\tilde c_{q,p}} }} $. On the other hand, if one wants to estimate each $|c_{q,p}|$ larger than a fixed $\delta
 $ with an uncertainty $\epsilon$, and obtain a correct value with probability $p$, the number of required repetitions is: $\tilde M_E \geq {k^2}/{4\delta^2\epsilon^2}$ where $k$ is chosen to satisfy
$ p=\erf{\frac{k}{\sqrt{2}}}$. Thus, the number of repetitions does not depend on $n$ but only on the precision $\epsilon$, the minumum measurable value $\delta$ and the probability of success $p$. The method is ``quantum efficient" as the number of quantum resources (i.e., copies of the quantum state, measurements, etc) is constant given a required precision. However, classical resources to determine every $c_{q,p}$ are still exponential in $n$ due to the fact that there are $N^2$ such coefficients. 

The setup of Figure 1 is a universal quantum state detector that, as opposed to previously proposed ones, is efficient. Using it to estimate a set of coefficient with fixed precision requires only a number of experiments (and classical resources) which is independent of $n$. Universal state detectors were introduced before \cite{2004EL.....65..165D} but, as we will show now, they are inefficient. These universal detectors do not use copies but ancillary systems prepared in a known state $\rho_0=\sum_{q,p} c^{(0)}_{q,p} T(q,p)/N$. When systems $(A)$ and $(B)$ are respectively prepared in states $\rho$ and $\rho_0$, it is simple to show that by performing joint Bell measurements on every pair of qubits we obtain 
\begin{equation}
c_{q,p} c^{(0)}_{q,p} = 
\left\langle T(q,p) \otimes T(q,p) \right\rangle_{\rho\otimes\rho_0}. \label{calphacalpha0}
\end{equation}
Therefore, knowing $c_{q,p}^{(0)}$ and measuring the expectation value appearing in \eqref{calphacalpha0} we can determine $c_{q,p}$. However, a universal detector must use a state $\rho_0$ that is not related with $\rho$. This is the origin of the inefficiency of the method, as can be seen as follows. Clearly, we can  determine $c_{q,p}$ only if the corresponding $c^{(0)}_{q,p}$ is non-vanishing. Moreover, the smaller the value of $c_{q,p}^{(0)}$ the higher the precision required in the estimation of $\left\langle T(q,p) \otimes T(q,p) \right\rangle_{\rho\otimes\rho_0}$. Consider first a state for which $\abs{c^{(0)}_{q,p}}$ are maximal. This is the case for stabilizer states (common eigenstates of a commuting set of $N$ Pauli operators $T(q,p)_S$). For such state there are $N$ non-vanishing coefficients $c^{(0)}_{{q,p}_S}$ taking values equal to $\pm 1$. For such $\rho_0$ the universal detector can only be used to estimate $N$ $c_{q,p}$'s providing no information about the 
 $N^2-N$ remaining ones, denying its universality. On the other hand all the coefficients could be estimated using a state $\rho_0$ with non-vanishing $c^{(0)}_{q,p}$ for all $(q,p)$'s. The problem for such unbiased $\rho_0$ is that all $c^{(0)}_{q,p}$'s are exponentially small. The reason for this is that $\sum_{q,p} {c^{(0)}}^2_{q,p} /N\le 1$. Therefore, each coefficient $c^{(0)}_{q,p}$ is $O(1/\sqrt{N})$. Then, if we use \eqref{calphacalpha0} to estimate them with fixed precision we need exponentially high precision in the estimation of the expectation value $\langle T(q,p) \otimes T(q,p)\rangle$.  For this reason the method is inefficient (the universal detector would have to be used an exponentially large number of times to achieve a fixed precision). Clearly, the use of a copy instead of an ancilla provides a simple way out of this problem. 

Full quantum state tomography is always exponentially hard as the number of unknown parameters scales as $4^n$. Therefore, it is crucial to conceive efficient methods for partial characterization of quantum states. Remarkably, the strategy described in Figure 1 is efficient also for this purpose. To see this we consider ``coarse grainded" Bell measurements: For any Bell state $|\beta_{m,n}\rangle$ we can estimate the probability to detect an even (odd) number of them in the measurement of all pairs. Then, we can compute $\Delta Prob_{m,n}=Prob({\rm even\ \#\ }|\beta_{m,n}\rangle) -
Prob({\rm odd\ \#\ }|\beta_{m,n}\rangle)$. It is simple to show that 
$\Delta Prob_{m,n}=Tr(O_{m,n}(\rho)\ \rho)$ where the (not necessarily positive) map $O_{m,n}$ is such that 
\begin{equation}
\Delta Prob_{m,n}= {1\over N}\sum_{q,p} s_{q,p} c_{q,p}^2. 
\label{evenodd-ij-2}
\end{equation}
Here, the $N^2$ components of the vector $s_{q,p}$ are $s_{q,p}=(-1)^{(m+1)(\alpha_x+\alpha_y)}(-1)^{(n+1)(\alpha_z+\alpha_y)}$, where $\alpha_x$ (resp.\!\! $\alpha_y$, $\alpha_z$) denotes the number of qubits for which the Pauli operator $T(q,p)$ contains an $X$ (resp.\!\! $Y$, $Z$) operator. For example, for the singlet $|\beta_{11}\rangle$, $s_{q,p} =1$ for every $\alpha$. For any other Bell state half of the components of $s_{q,p}$ are equal to $+1$ and the other half are equal to $-1$. For the singlet the above formula reduces to 
\begin{equation}
\Delta Prob_{1,1}={1\over N}\sum_{q,p} c_{q,p}^2=\Tr{\rho^2}.
\label{evenodd-singlet}
\end{equation}
Thus, this measurement reveals the purity of the state. Partial purities can be detected in the same way: Consider the state $\rho_J$, obtained after tracing out the qubits for which the binary $n$--tuple $J$ is zero. Purity  of such state is the sum of $c_{q,p}^2$ for the coefficients associated with Pauli operators containing the identity in the qubits for which the corresponding component of $J$ is zero. To obtain it we must use \eqref{evenodd-singlet} counting singlets only in the qubits where the corresponding bit of $J$ is equal to unity. 

The above method for estimating purity is equivalent to the one proposed by Ekert {\it{et al}} who used the fact that purity is equal to the expectation value of the swap operator in the state $\rho^{(A)}\otimes\rho^{(B)}$ \cite{2002PhRvL..88u7901E}. But our results also show that by making more general coarse grained Bell measurements we are not only able to efficiently detect partial purities. In fact, with the same effort we reveal other quantities that partially characterize the quantum state and have the form $\Delta P_S=\left(\sum_{(q,p)\in S}c_{q,p}^2- \sum_{(q,p)\in\bar S} c_{q,p}^2\right)/N$,  where $\DE{S,\bar S}$ is a partition of the $N^2$ coefficients $c_{q,p}$ in two halves. It is possible to generalize this even further by grouping Bell states in each pair of qubits (in this case one can attain linear combinations with vectors $s_{q,p}$ that have a different number of $\pm 1$ components). Other wheighted  sums of squares of $c_\alpha$ over certain sets of $(q,p)$'s can also be obtained in this way. 

It is interesting to consider another related coarse grained Bell measurement: we can estimate the probability $p_{m,n}^{(all)}$ to find all pairs of qubits in the subspace orthogonal to $|\beta_{m,n}\rangle$. For the case of the singlet this is related to the multipartite concurrence that for a pure $n$--qubit state is \cite{2004PhRvL..93w0501C}
\begin{equation}
 \mathcal{C}\left( \rho\right)=2^{1-n/2}\sqrt{\left(2^n-2\right)-\sum_l \Tr{\rho_l^2}}. 
\end{equation}
Here the $n$--tuple $l$ labels every nontrivial subset of the $n$ qubits. As noticed in \cite{2006PhRvL..97e0501A}, this can be rewritten as:
\begin{equation}
 \mathcal{C}\left( \rho \right)=2\sqrt{1-p_{1,1}^{(all)} },
\end{equation}
therefore, concurrence of a pure state can be estimated efficiently using the type of coarse grained Bell measurements we described above. The probability $p_{1,1}^{(all)}$ is also a quadratic form of the coefficients $c_{q,p}$: $p_{1,1}^{(all)} =\sum_{q,p} c_{q,p}^2 3^{\alpha_0}/N^2$, where $\alpha_0$ is the number of qubits for which $T(q,p)$ contains a factor equal to the identity. More generally, if we measure $ p_{m,n}^{(all)}$ we obtain a quantity that is not a concurrence but provides different tomographic information. It can be expressed as $ p_{m,n}^{(all)} =\sum_{q,p} c_{q,p}^2 f_{q,p}/N^2$, where $f_{q,p} =3^{\alpha_0}(-1)^{(m+1)(\alpha_x+\alpha_y)+(n+1)(\alpha_z+\alpha_y)}$. 

As we mentioned, for an efficient partial QST it is necessary to estimate certain differences between probabilities such as in (3). Such probabilities, as the ccPMVM Theorem states, are fidelities of ccP maps. Thus, the right hand side of equation (3) can be expressed as the fidelity of a map which is the difference between two ccP maps: $\Delta P_{m,n}=Tr(\rho O_{m,n}(\rho))$ ($O_{m,n}$ is not ccP). One of such maps is the identity, which is not ccP since when composed with the transposition becomes non-cP. For that case, one finds $\Delta P_{1,1}=Tr(\rho^2)$. Thus, purity can never be evaluated as the probability of a generalized measurement but only as the difference between probabilities. In turn, fidelities of positive maps that are not ccP can be related via the Jamio\l kowski isomorphism with expectation values of Hermitian operators (not necessarily positive). Such Hermitian operator can be written as the difference between two positive operators. For the example disc
 ussed above, where the positive map is the identity, it turns out that the non-positive operator is the one implementing the swap between the two copies as the well known identity $\Delta P_{1,1}=Tr(\rho^2)=Tr(Swap\ \rho\otimes\rho)$ shows. Summarizing, we showed that by performing general measurements on two copies of a quantum state one always detects probabilities that are fidelities of completely co-positive maps. We also described techniques to perform efficient quantum state tomography by taking advantage of measurements with copies.
 
JPP is a fellow of CONICET. This work was supported with grants from ANPCyT, UBACyT, Fapemig, and Santa Fe Institute. MTC visit to UBA was supported by an AUGM collaboration grant.

\end{document}